\documentclass[a4paper,conference]{llncs}

\usepackage{xspace}
\usepackage{tikz}
\usetikzlibrary{shapes,arrows,positioning,decorations.markings,chains,fit,shapes.multipart}
\usepackage{import}
\usepackage{times}
\usepackage{graphicx}
\usepackage{cite}
\usepackage{amsfonts}
\usepackage{amssymb}
\usepackage{amsmath}
\usepackage{mathptm}
\usepackage{listings}
\usepackage{verbatim}
\usepackage{alltt}
\usepackage{mdwlist}
\usepackage{hyperref}

\newcommand{\tool}[1]{\textsc{#1}\xspace}
\newcommand{\slec}{\tool{slec}}
\newcommand{\hwcbmcv}{\tool{hw-cbmc}}
\newcommand{\hwcbmcver}{\tool{hw-cbmc 5.4}}
\newcommand{\verifoxver}{\tool{VerifOx 0.1}}
\newcommand{\verifox}{\tool{VerifOx}}
\newcommand{\klee}{\tool{klee}}
\begin{document}

\title{Equivalence Checking a Floating-point Unit\\
       against a High-level C Model\\ 
       \vskip3mm \large Extended Version}

\author{Rajdeep Mukherjee\inst{1} \and
Saurabh Joshi\inst{2} \and
Andreas Griesmayer\inst{3} \and \\
Daniel Kroening\inst{1} \and 
Tom Melham\inst{1}}

\institute{University of Oxford, UK \and IIT Hyderabad, India \and ARM Limited \\
\email{\{rajdeep.mukherjee,kroening,tom.melham\}@cs.ox.ac.uk},
\email{sbjoshi@iith.ac.in},~\email{andreas.griesmayer@arm.com}}

\maketitle

\begin{abstract}
Semiconductor companies have increasingly adopted a methodology that 
starts with a system-level design specification in C/C++/SystemC. This 
model is extensively simulated to ensure correct functionality and
performance.  Later, a Register Transfer Level (RTL) implementation is
created in Verilog, either manually by a designer or 
automatically by a high-level synthesis tool. It is essential to check 
that the C and Verilog programs are consistent. In this paper, we present 
a two-step approach, embodied in two equivalence checking tools, \verifox 
and \hwcbmcv, to validate designs at the software and RTL levels, respectively. 
\verifox is used for equivalence checking of an untimed software model in 
C against a high-level reference model in C. 
\hwcbmcv verifies the equivalence of a Verilog RTL
implementation against an untimed software model in C. To evaluate our
tools, we applied them to a commercial floating-point arithmetic unit (FPU) 
from ARM and an open-source dual-path floating-point adder.
\end{abstract}            

\section{Introduction}

\begin{figure}[tb]
\scalebox{.55}{\import{figures/}{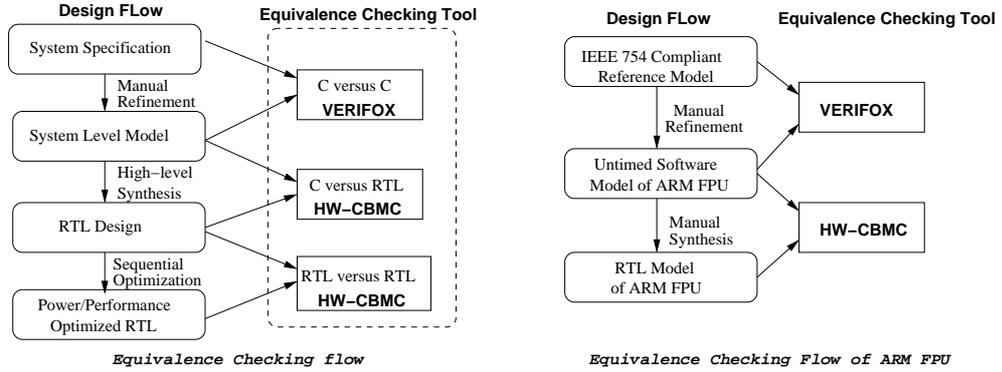}}
\caption{Electronic Design Automation Flow 
\label{fig:eda-flow}}
\end{figure}

One of the most important tasks in Electronic Design Automation (EDA) 
is to check whether the low-level implementation (RTL or gate-level) 
complies with the system-level specification. 
Figure~\ref{fig:eda-flow} illustrates the role of equivalence checking 
(EC) in the design process. In this paper, we present 
a new EC tool, \verifox, that is used for equivalence checking of an 
untimed software (SW) model against a high-level reference model. 
Later, a Register Transfer Level (RTL) model is implemented, 
either manually by a hardware designer or 
automatically by a synthesis tool. To guarantee that 
the RTL is consistent with the SW model, we use an existing tool,
\hwcbmcv~\cite{CKY03}, to check the correctness of the synthesized hardware 
RTL against a SW model.

In this paper, we address the most general and thus most difficult variant
of EC: the case where the high-level and the low-level design are
substantially different. State-of-the-art tools, such as
Hector~\cite{DBLP:conf/date/KoelblJJP09} from Synopsys and \slec from
Calypto,\footnote{http://calypto.com/en/products/slec/} rely on
\emph{equivalence points}~\cite{DBLP:conf/dac/WuH06}, and hence they are
ineffective in this scenario. We present an approach based on bounded
analysis, embodied in the tools \verifox and \hwcbmcv, that can handle
arbitrary designs.  

\verifox is used for equivalence checking of an untimed
software model against a high-level reference model and \hwcbmcv is used for
equivalence checking of the RTL implementation against a software model.
EC is broadly classified into two separate categories: combinational
equivalence checking (CEC) and sequential equivalence checking (SEC). 
CEC~is used for a pair of models that are cycle accurate and have the same
state-holding elements. SEC is used when the
high-level model is not cycle accurate or has a substantially different set
of state-holding elements~\cite{DBLP:conf/date/Eijk98,
DBLP:conf/iccd/BaumgartnerMPKJ06}. It is well-known that EC of 
floating-point designs is
difficult~\cite{DBLP:conf/aspdac/XueCS13,DBLP:conf/cav/Fujita96}. So
there is a need for automatic tools that formally validate floating-point
designs at various stages of the synthesis flow, as illustrated by right side
flow of Figure~\ref{fig:eda-flow}.

\medskip \noindent \textbf{\emph{Contributions: }}
In this paper, we sketch two significant equivalence-verification tools:
\begin{enumerate}
\item {\verifox, a tool for equivalence checking of software models 
given as C programs.} We present a path-based symbolic 
execution tool, \verifox, for bounded equivalence checking 
of floating-point software implementations against a IEEE 
754 compliant reference model. \verifox supports C89, 
C99 standards in the front-end. \verifox also supports 
SAT and SMT backends for constraint solving. \verifox is available 
at~\url{http://www.cprover.org/verifox}.
\item {\hwcbmcv, a tool for C versus RTL equivalence checking.}
\hwcbmcv is used for bounded equivalence checking of Verilog RTL 
against C/C++ models. \hwcbmcv supports IEEE 1364-2005 System Verilog 
standards and the C89, C99 standards. \hwcbmcv generates a monolithic 
formula from the C and RTL description, which is then checked with 
SAT/SMT solvers. 
\end{enumerate} 

\section{\verifox: A tool for equivalence checking of C programs} 
\verifox is a path-based symbolic execution tool for equivalence checking of
C programs. The tool architecture is shown on the left side of
Figure~\ref{fig:tools}. \verifox supports the C89 and C99 standards. The
key feature is symbolic reasoning about equivalence between FP
operations. To this end, \verifox implements a model of the core IEEE
754 arithmetic operations---single- and double-precision addition,
subtraction, multiplication, and division---which can be used as reference
designs for equivalence checking.
So \verifox does not require external reference models for
equivalence checking of floating-point designs. This significantly
simplifies the users effort to do equivalence checking at
software level. The reference model in \verifox is equivalent to the
Softfloat
model.\footnote{http://www.jhauser.us/arithmetic/SoftFloat.html}
\verifox also supports SAT and SMT backends for constraint
solving.

\begin{figure}[tb]
\scalebox{.55}{\import{figures/}{tools.pspdftex}}
\caption{\verifox and \hwcbmcv Tool Architecture
\label{fig:tools}}
\end{figure}

\begin{figure}[b]
\scriptsize
\begin{tabular}{l|l|l|l|l}
\hline
Program & Path & Path & Path & Monolithic\\
        & Constraint 1 & Constraint 2 & Constraint 3 & Path Constraint\\
\hline
\begin{lstlisting}[mathescape=true,language=C]
void top(){
 if(reset) {
   x=0;
   y=0; }
 else {
  if(a > b)
   x=a+b;
  else
   y=(a & 3)<<b; }}
\end{lstlisting}
&
\begin{minipage}{1.5cm}
$\begin{array}[t]{@{}l}
C_1 \; \equiv \\
\;\;\mathit{\mathit{reset}}_1 \neq 0\; \land \\
\;\; x_2 = 0\;\land \\
\;\; y_2 = 0
\end{array}$
\end{minipage}
&
\begin{minipage}{1.5cm}
$\begin{array}[t]{@{}l}
C_2 \; \equiv \\
\;\;\mathit{\mathit{reset}}_1=0\; \land \\
\;\;b_1 \ngeq a_1\; \land \\
\;\;x_3 = a_1 + b_1
\end{array}$
\end{minipage}
&
\begin{minipage}{2cm}
$\begin{array}[t]{@{}l}
C_3\; \equiv \\
\;\;\mathit{\mathit{reset}}_1=0\; \land \\
\;\;b_1 \ge a_1\; \land \\
\;\;y_3 = (a_1\&3) \\
\;\;<\!\!<b_1
\end{array}$
\end{minipage}
&
\begin{minipage}{3.75cm}
$\begin{array}{l@{\,\,}c@{\,\,}l}
C &\iff& ((\mathit{\mathit{guard}}_1 =  \neg(\mathit{\mathit{reset}}_1 = 0)) \land \\
  &    & (x_2 = 0) \land (y_2 = 0) \land \\
  &    & (x_3 = x_1) \land (y_3 = y_1) \land \\
  &    & (\mathit{guard}_2 = \neg(b_1 >= a_1)) \land \\
  &    & (x_4 = a_1 + b_1) \land (x_5 = x_3) \land \\
  &    & (y_4 = (a_1 \& 3) <\!\!< b_1) \land \\
  &    & (x_6 = \mathit{ite}(\mathit{\mathit{guard}}_2, x_4, x_5)) \land \\
  &    & (y_5 = \mathit{ite}(\mathit{\mathit{guard}}_2, y_3, y_4)) \land \\
  &    & (x_7 = \mathit{ite}(\mathit{\mathit{guard}}_1, 0, x_6)) \land \\
  &    & (y_6 = \mathit{ite}(\mathit{\mathit{guard}}_1, 0, y_5)))
\end{array}$
\end{minipage}
\\
\hline 
\end{tabular} 
\caption{Single-path and Monolithic Symbolic Execution}
\label{figure:simulation}
\end{figure}

Given a reference model, an implementation model in C and a set of partition
constraints, \verifox performs depth-first exploration of program paths with
certain optimizations, such as eager infeasible path pruning and incremental
constraint solving. This enables automatic decomposition of the verification 
state-space into subproblems, by input-space and/or state-space decomposition. 
The decomposition is done in tandem in both models, exploiting the structure 
present in the high-level model. The approach generates many but simpler SAT/SMT 
queries, similar to the technique followed in \klee~\cite{DBLP:conf/osdi/CadarDE08}.
The main focus of our technique is to pass only those verification
conditions to the underlying solver for which the corresponding path conditions
are feasible with respect to the property under consideration and the 
partitioning constraints such as case splitting. 

Figure~\ref{figure:simulation} shows three feasible path 
constraints corresponding to the three paths in the program on the
left. In contrast, the last column of Figure~\ref{figure:simulation} shows
monolithic path-constraint generated by \hwcbmcv.

\medskip \noindent \textbf{\emph{Incremental solving in \verifox.}} \verifox can be run in two 
different modes: partial incremental and 
full incremental. In partial incremental mode, only one solver instance is maintained 
while going down a single path. So when making a feasibility
check from one branch $b_1$ to another branch $b_2$ along a single path, only the program
segment from $b_1$ to $b_2$ is encoded as a constraint and added to the existing solver
instance. Internal solver states and the information that the solver gathers during the search
remain valid as long as all the queries that are posed to the solver in succession are
monotonically stronger. If the solver solves a formula $\phi$, then posing $\phi \wedge \psi$ as 
a query to the same solver instance allows one to reuse solver knowledge 
it has already acquired, because any assignment that falsifies $\phi$ also
falsifies $\phi \wedge \psi$. Thus the solver need not revisit the assignments that 
it has already ruled out. This results in speeding up the feasibility 
check of the symbolic state at $b_2$, as the feasibility check at $b_1$ was $\mathit{true}$. 
A new solver instance is used to explore a different path, after the current path is 
detected as infeasible.

In full incremental mode, only one solver instance is maintained throughout the whole 
symbolic execution. Let $\phi_{b_1b_2}$ denote the encoding of the path fragment
from $b_1$ to $b_2$. It is added in the solver as $B_{b_1b_2} \Rightarrow \phi_{b_1b_2}$.
Then, $B_{b_1b_2}$ is added as a \textit{blocking variable}\footnote{The SAT community uses the
term \textit{assumption variables} or \textit{assumptions}, but we will use the 
term blocking variable to avoid ambiguity with assumptions in the program.}
to enforce constraints specified by $\phi_{b_1b_2}$. Blocking variables are 
treated specially inside the solvers: unlike regular variables or clauses, 
the blocking can be removed in subsequent queries without invalidating the 
solver instance. When one wants to back-track the symbolic execution, the 
blocking $B_{b_1b_2}$ is removed and a unit clause $\neg B_{b_1b_2}$ is 
added to the solver, thus effectively removing $\phi_{b_1b_2}$.

\section{\hwcbmcv: A tool for equivalence checking of C and RTL} 

\hwcbmcv is used for bounded equivalence checking of C and 
Verilog RTL. The tool architecture is shown on the right side 
of Figure~\ref{fig:tools}. \hwcbmcv supports IEEE 1364-2005 System Verilog 
standards and the C89, C99 standards. \hwcbmcv maintains two 
separate flows for hardware and software. The top flow in 
Figure~\ref{fig:tools} uses synthesis to obtain either a bit-level 
or a word-level netlist from Verilog RTL. The bottom flow 
illustrates the translation of the C program into static single 
assignment (SSA) form~\cite{Cytron:1989:EMC:75277.75280}.
These two flows meet only at the solver. 
Thus, \hwcbmcv generates a monolithic formula from the C and RTL 
description, which is then checked with SAT/SMT solvers. \hwcbmcv 
provides specific handshake primitives such as $next\_timeframe()$ 
and $set\_inputs()$ that direct the tool to set the inputs to the 
hardware signals and advance the clock, respectively. The details of \hwcbmcv are
available online.\footnote{http://www.cprover.org/hardware/sequential-equivalence/}

\section{Experimental Results}

In this section, we report experimental results for equivalence 
checking of difficult floating-point designs. All our 
experiments were performed on an Intel\textsuperscript{\textregistered} Xeon\textsuperscript{\textregistered} machine with 3.07\,GHz clock speed
and 48\,GB RAM.  All times reported are in seconds. 
MiniSAT-2.2.0~\cite{DBLP:conf/sat/EenB05} was used as underlying 
SAT solver with \verifoxver and \hwcbmcver. The timeout 
for all our experiments was set to 2~hours.

\medskip\noindent\textbf{\emph{Proprietary Floating-point Arithmetic Core: }} 
We verified parts of a floating-point arithmetic unit (FPU) of a next
generation ARM\textsuperscript{\textregistered} GPU. The FP core is primarily composed of single-
and double-precision {\em ADD}, {\em SUB}, {\em FMA} 
and {\em TBL} functional units, the register files,
and interface logic. The pipelined computation unit implements 
FP operations on a 128-bit data-path. In this paper, we 
verified the single-precision addition ({\em FP-ADD}), 
rounding ({\em FP-ROUND}), minimum ({\em FP-MIN}) 
and maximum ({\em FP-MAX}) operations. The FP-ADD unit 
can perform two operations in parallel by using two 64-bit
adders over multiple pipeline stages.
Each 64-bit unit can also perform operations with 
smaller bit widths. The FPU decodes the incoming 
instruction, applies the input modifiers and 
provides properly modified input data to the respective 
sub-unit. The implementation is around~38000 LOC, 
generating tens of thousands of gates. We obtained the SW 
model (in C) and the Verilog RTL model of the FPU core from ARM.
(Due to proprietary nature of the FPU design, we can not 
share the commercial ARM IP.)

\medskip\noindent\textbf{\emph{Open-source Dual-path Floating-point Adder: }}
We have developed both a C and a Verilog implementation of an
\mbox{IEEE-754} 32-bit single-precision dual-path floating point
adder/subtractor.  This floating-point
design includes various modules for packing, unpacking, normalizing,
rounding and handling of infinite, normal, subnormal, zero and NaN
(Not-a-Number) cases. We distribute the C and RTL implementation of 
the dual-path FP adder at~\url{http://www.cprover.org/verifox}. 

\medskip\noindent\textbf{\emph{Reference Model: }} 
The IEEE 754 compliant floating-point implementations 
in \verifox are used as the golden reference model 
for equivalence checking at the software level. For 
equivalence checking at the RTL phase, we used the untimed software model 
from ARM as the reference model, as shown on the right side of 
Figure~\ref{fig:eda-flow}. 

\medskip\noindent\textbf{\emph{Miters for Equivalence Checking: }}
\lstset{language=C,basicstyle=\ttfamily}
A miter circuit~\cite{Brand} is built from two given circuits $A$ and $B$ as follows:
identical inputs are fed into $A$ and $B$, and the outputs of $A$ and $B$
are compared using a comparator. For equivalence checking at software level, 
one of the circuits is a SW program and the other is a high-level 
reference model. For the RTL phase, one of the circuits is a 
SW program treated as reference model and the other is an RTL implementation.

\medskip\noindent\textbf{\emph{Case-splitting for Equivalence Checking: }}
Case-splitting is a common practice to scale up formal 
verification~\cite{DBLP:conf/cav/Fujita96, DBLP:conf/aspdac/XueCS13,
DBLP:conf/date/KoelblJJP09} and is often performed 
by user-specified assumptions. The \texttt{CPROVER\_assume(c)} 
statement instructs \hwcbmcv and \verifox to restrict the analysis 
to only those paths satisfying a given condition \texttt{c}. For 
example, we can limit the analysis to those paths that are exercised 
by inputs where the rounding mode is nearest-even (RNE) and both 
input numbers are NaNs by adding the following line:

\medskip

\noindent\quad\texttt{CPROVER\_assume(roundingMode==RNE \&\& uf\_nan \&\& ug\_nan);}
%\texttt{CPROVER\_assume(fp\_add\_sub.uf\_nan);} \\
%\texttt{CPROVER\_assume(fp\_add\_sub.ug\_nan);}

\medskip\noindent\textbf{\emph{Discussion of Results: }} 
Table~\ref{table:result} reports the run times for equivalence 
checking of the  ARM FPU and the dual-path FP adder. Column~1 
gives the name of FP design and columns~2--6 show the runtimes 
for partition modes INF, ZERO, NaN, SUBNORMAL, and NORMAL respectively. 
For example, the partition constraint `INF' means addition of two 
infinite numbers. Column~7 reports the total time for equivalence 
checking without any partitioning. 

\verifox successfully proved the 
equivalence of all FP operations in the SW implementation of ARM FPU 
against the built-in reference model. However, a bug in FP-MIN and 
FP-MAX (reported as ERROR in Table~\ref{table:result}) is detected by 
\hwcbmcv in the RTL implementation of ARM FPU when checked 
against the SW model of ARM FPU for the case when both the input 
numbers are NaN.  This happens mostly due to bugs in the high-level 
synthesis tool or during manual translation of SW model to RTL. 
\verifox and \hwcbmcv is able to detect bugs in the SW 
and RTL models of these designs respectively -- thereby 
emphasizing the need for equivalence checking to validate 
the synthesis process during the EDA flow.
Further, we investigate the reason for higher verification times for 
subnormal numbers compared to normal, infinity, NaN's 
and zero's. This is attributed to higher number of paths in 
subnormal case compared to INF, NaN's and zero's. Closest to 
our floating-point symbolic execution technique in \verifox 
is the tool \textsc{KLEE-FP}~\cite{DBLP:conf/eurosys/CollingbourneCK11}. 
We could not, however, run \textsc{KLEE-FP} on the software models because 
the front-end of KLEE-FP failed to parse the ARM models. 

\begin{table}[t]
\begin{center}
{
\begin{scriptsize}
\begin{tabular}{|l|l|l|l|l|l|l|}
\hline
  &  \multicolumn{5}{c|}{Case-splitting} & \multicolumn{1}{c|}{No-partition} \\ 
\cline{2-7}
  Design & INF & ZERO & NaN & SUBNORMAL & NORMAL & Total    \\
  \cline{1-7} 
 \multicolumn{7}{|c|}{Equivalence checking at Software Level (\verifox)} \\
 \hline
FP-ADD & 9.56 & 11.54 & 9.95 & 1124.18 & 77.74 & 1566.72 \\ \hline 
FP-ROUND & 1.24 & 1.36 & 1.32 & 3.78 & 1.63 & 4.71 \\ \hline 
FP-MIN & 9.76 & 9.85 & 9.78 & 28.67 & 9.86 & 48.70 \\ \hline 
FP-MAX & 9.80 & 9.88 & 9.97 & 28.70 & 9.90 & 35.81 \\ \hline
DUAL-PATH ADDER & 3.15 & 3.11 & 2.14 & 88.12 & 55.28 & 497.67 \\ \hline 
\multicolumn{7}{|c|}{Equivalence checking at RTL (\hwcbmcv)} \\ \hline
FP-ADD & 18.12 & 18.02 & 17.87 & 18.73 & 39.60 & 40.72 \\ \hline 
FP-ROUND & 11.87 & 12.73 & 13.44 & 13.67 & 14.03 & 14.11 \\ \hline 
FP-MIN & 13.72 & 13.62 & ERROR & 14.10 & 14.08 & 14.15  \\ \hline 
FP-MAX & 13.70 & 13.58 & ERROR & 14.09 & 14.06 & 14.05 \\ \hline 
DUAL-PATH ADDER & 0.88 & 0.87 & 0.99 & 169.49 & 22.42 & 668.61 \\ \hline 
\end{tabular}
\end{scriptsize}
}
\end{center}
\vspace{-1.3mm}
\caption{Equivalence checking of ARM FPU and DUAL-PATH Adder (All time in seconds)
\label{table:result}}
\end{table}

\section{Related work}

The concept of symbolic execution~\cite{DBLP:journals/tse/Clarke76,
DBLP:conf/pldi/GodefroidKS05,DBLP:conf/osdi/CadarDE08} is prevalent in the
software domain for automated test generation as well as bug finding.  
Tools such as Dart~\cite{DBLP:conf/pldi/GodefroidKS05}, Klee~\cite{DBLP:conf/osdi/CadarDE08},
EXE~\cite{DBLP:journals/tissec/CadarGPDE08}, 
Cloud9~\cite{DBLP:conf/pldi/KuznetsovKBC12} employ such a technique for
efficient test case generation and bug finding. By contrast, we used 
path-wise symbolic execution for equivalence checking of software models
against a reference model. A user-provided assumption specifies certain testability 
criteria that render majority of the design logic 
irrelevant~\cite{DBLP:conf/cav/Fujita96, DBLP:conf/aspdac/XueCS13,
DBLP:conf/date/KoelblJJP09}, thus giving rise to 
large number of infeasible paths in the design.  
Conventional SAT-based bounded model checking~\cite{biere,CKY03, 
Clarke:2003:HVU:1119772.1119831} can not exploit this infeasibility 
because these techniques create a monolithic formula by unrolling 
the entire transition system up to a given bound, which is then 
passed to SAT/SMT solver. These tools perform case-splitting at 
the level of solver through the effect of constant propagation. 
Optimizations such as eager path pruning 
combined with incremental encoding enable \verifox to 
address this limitation. 

\section{Concluding Remarks}

In this paper we presented \verifox, our path-based symbolic execution
tool, which is used for equivalence checking of arbitrary software
models in C. The key feature of \verifox is symbolic reasoning on 
the equivalence between floating-point operations. To this end, 
\verifox implements a model of the core IEEE 754 arithmetic 
operations, which can be used for reference models.
Further, to validate the synthesis of RTL from software model, we used 
our existing tool, \hwcbmcv, for equivalence checking of RTL designs 
against the software model used as reference. We successfully 
demonstrated the utility of our equivalence checking tool chain, 
\verifox and \hwcbmcv, on a large commercial FPU core from ARM and 
a dual-path FP adder. Experience suggests that the synthesis of 
software models to RTL is often error prone---this emphasizes
the need for automated equivalence checking tools at various 
stages of EDA flow. In the future, we plan to investigate 
various path exploration strategies and path-merging techniques 
in \verifox to further scale equivalence checking to complex 
data and control intensive designs.   

\section*{Acknowledgements}

Part of the presented work was conducted during an internship at ARM.
The authors want to thank in particular Luka Dejanovic, Joe Tapply,
and Ian Clifford for their help with setting up the experiments.

% =====================================================================
% References
% =====================================================================

\bibliographystyle{splncs03}
\bibliography{paper} 

% =====================================================================
% Appendix 
% =====================================================================

\newpage
\appendix
\section{Appendix}

This appendix provides simple, illustrative examples of the use of \verifox and \hwcbmcv, as well as further technical details.

\renewcommand{\topfraction}{0.9}	  % 0.8 of the top page can be fig.
\renewcommand{\bottomfraction}{0.9}	  % 0.8 of the bottom page can be fig.
\renewcommand{\textfraction}{0.0}	  % 0.0 of the page must contain text

\subsection*{Worked Example of \verifox}

\lstdefinestyle{base}{
  language=C,
  emptylines=1,
  breaklines=true,
  basicstyle=\ttfamily\color{black},
  moredelim=**[is][\color{red}]{@}{@},
}

Figure~\ref{fig:example} demonstrates the working of \verifox as a 
property verifier in the absence of a reference model. Note that 
equivalence checking is a special case of property verification 
where the property is replaced by a reference model. Hence, \verifox 
can be configured as a property verifier or as an equivalence checker.

Let us consider a software model as shown in column~1 
in Figure~\ref{fig:example}. The program implements a high-level 
power management strategy to orchestrate various modules, such as, 
{\em core}, {\em memory} etc. Depending on the interrupt
status ($env$), power modes ($mode$) and power gated logic ($power\_gated$),
the call to {\em core} or {\em memory} is made. These units are
complex implementations of a processor core or a memory unit.

\begin{figure}[!b]
\scriptsize
\begin{tabular}{l|l|l}
\hline
Software Model (in C) & Sliced program (wrt.
property) & Pruned program (wrt. assumptions) \\
\hline
\begin{lstlisting}[mathescape=true,language=C]
#define threshold 15
if(reset) {
  mode=TURN_OFF;
  feedback=0; 
 } else { 
 // Trigger IP if env is set
 if(env) { 
 // check the voltage level
 if(voltage_level < threshold)
  power_gated = 1;
 else power_gated = 0;
 // check the low-power modes
 if(mode == STAND_BY || 
   mode == TURN_OFF) {
 // power gated logic, 
 // call to core  
 if(power_gated) {
  core(reset,mode,power_gated
  ser_in,&buf_out);
  feedback = LOW;
 }
 else { // normal logic
   core(reset,mode,power_gated
   ser_in,&buf_out);
   feedback = buf_out; }} 
 }
 else 
  // call to memory 
  memory(size);
 }
\end{lstlisting}
&
\begin{lstlisting}[mathescape=true,language=C]
#define threshold 15
if(reset) {
 mode=TURN_OFF;
 feedback=0;
 } else { 
 // Trigger IP if env is set
 if(env) {
 // check the voltage level
 if(voltage_level < threshold)
  power_gated = 1;
 else power_gated = 0;
 // check the low-power modes
 if(mode == STAND_BY || 
 mode == TURN_OFF) {
 // power gated logic, 
 // call to core  
 if(power_gated) {
  core(reset,mode,power_gated
  ser_in,&buf_out);
  feedback = LOW;
 }
 else { // normal logic
  core(reset,mode,power_gated
  ser_in,&buf_out);
  feedback = buf_out;
  }
 }} 
}
$\begin{array}{l@{}c@{}l}
Assertion: (!(rst!=0) || (feedback>0))
\end{array}$
\end{lstlisting}
&
\begin{lstlisting}[mathescape=true,language=C]
$\begin{array}{l@{}c@{}l}
Assumption:(reset == 0) \wedge (env == 1)  \\
\wedge (mode == STAND\_BY) \\
\wedge (voltage\_level == 10)
\end{array}$

#define threshold 15
// non-reset logic
else {
if(env) {
 // check the voltage level
 if(voltage_level < threshold)
  power_gated = 1;
 
 // check the low-power modes
 if(mode == STAND_BY || 
   mode == TURN_OFF) {
  // power gated logic, 
  // call to core  
  if(power_gated) {
   core(reset,mode,power_gated
   ser_in,&buf_out);
   feedback = LOW;
   }
  }
 }
}
$\begin{array}{l@{}c@{}l}
Assertion: (!(rst!=0) || (feedback>0))
\end{array}$
\end{lstlisting}
\\
\hline
\end{tabular}
\caption{Example demonstrating automated slicing and path pruning in \verifox}
\label{fig:example}
\end{figure}

State-of-the-art verification tools may not be able verify the whole system
due to resource limitations. Therefore, it is a common practice to write
additional constraints, also known as {\em assumptions}, that exercise only
a fragment of the entire state-space. Verification engine can use these
assumptions to partition the state-space, thus decomposing a hard proof
into simpler sub-proofs.

Column~2 presents the result of property-driven
slicing on the input program. This step is purely syntactic, meaning that
we perform a backward dependency analysis~\cite{icse81} starting from the
property which only preserve those program fragments that are relevant to
the given property. The sliced program is then passed to the symbolic
execution engine that performs eager infeasibility based path-pruning. The
result of infeasible path pruning based on assumption is shown in column~3. 
This step is semantic because \verifox determines the feasibility of paths
in the sliced program in an eager fashion with respect to the user-provided
assumptions using satisfiability queries.  

An important point to note here
is that the number of path constraints after slicing and infeasible path
pruning are significantly less compared to the initial program. 
Additionally, these per-path constraints are much easier to solve compared
to a monolithic formula generated from a BMC-style symbolic execution tool.

\medskip\noindent\textbf{\emph{Command to run \verifox.}}
Below are the commands to run \verifox in partial or full incremental mode.
When \verifox is used as an equivalence checker, the input file is  usually a 
miter in C which must include both the reference model and the implementation model.
However, in the absence of a reference model, one can write assertions inside the 
software model to configure \verifox as a property verifier. The command line 
switch \texttt{--unwind} is used to specify the unwind depth for the software model.
To use the SMT backend with \verifox, the command line switch is
\texttt{--smt2}, followed by the name of the SMT solver, for example
\texttt{--z3}. Note that the SMT solver must be installed in the 
system. The switch \texttt{--help} shows the available command line options 
for using \verifox.  

\begin{small}
\begin{verbatim}
// partial incremental mode with SAT
verifox-pi filename.c --unwind N 
// full incremental mode with SAT
verifox-fi filename.c --unwind N
// partial incremental mode with SMT
verifox-pi filename.c --smt2 --z3
\end{verbatim}
\end{small}

\begin{figure}[t]
\scriptsize
\begin{tabular}{l|l|l}
\hline
C Program & Verilog RTL & Miter \\
\hline
\begin{lstlisting}[mathescape=true,language=C]
struct st_up_counter{
  unsigned char out;
};
struct st_up_counter 
sup_counter;

void upcounter(unsigned char *out, 
  _Bool enable, _Bool clk, 
  _Bool reset)
{
 unsigned char out_old;
 out_old = sup_counter.out;
 if(reset)
 {
  sup_counter.out = 0;
 }
 else if(enable)
 {
  sup_counter.out = 
   out_old + 1;
 }
}
\end{lstlisting}
&
\begin{lstlisting}[mathescape=true,language=C]
module up_counter(out, 
    enable,clk, reset);
output [7:0] out;
input enable, clk, reset;
reg [7:0] out;
always @(posedge clk)
if(reset) 
begin
 out<=8'b0;
end 
else if(enable) 
begin
out<=out+1;
end
endmodule 
\end{lstlisting}
& 
\begin{lstlisting}[mathescape=true,language=C]
typedef unsigned char _u8;
struct module_up_counter {
  _u8 out;
  _Bool enable;
  _Bool clk;
  _Bool reset;
};
extern struct 
module_up_counter up_counter;
int main()
{
  // Inputs of C program
  _Bool enable;
  _Bool clk;
  _Bool reset;
  unsigned char out;
  
	// reset the design
  // call to C function
  upcounter(&out, 0, clk, 1);
  // set Verilog inputs
  up_counter.enable = 0;
  up_counter.reset = 1;
  set_inputs();
  next_timeframe();
  assert(up_counter.out 
    == sup_counter.out);
  
  while(1) {	
   // Start counting, set 
   // enable = 1 and reset = 0
   up_counter.reset = 0;
   up_counter.enable = 1;
   set_inputs();
   next_timeframe();
   upcounter(&out, 1, clk, 0);
   assert(up_counter.out 
    == sup_counter.out);
  }
}  
\end{lstlisting}
\\
\hline
\end{tabular}
\caption{Example of equivalence checking using \hwcbmcv}
\label{fig:hwcbmc-example}
\end{figure}

\subsection*{Worked Example of \hwcbmcv}

Figure~\ref{fig:hwcbmc-example} demonstrates the working of \hwcbmcv as a 
C-RTL equivalence checker. Columns~1--3 present a C model of an
up-counter, an RTL model of the same device, and a miter that feeds the same input to 
the C and RTL model and asserts equivalence of their outputs. 
\hwcbmcv can be configured in {\em bit-level} or {\em word-level} mode. In bit-level 
mode, the input models are synthesized to 
And Inverter Graphs (AIG)\footnote{http://fmv.jku.at/aiger/}  
and then passed to the SAT solver. In word-level mode, 
the input models are synthesized into an intermediate word-level format, which are then 
despatched to a word-level SMT solver.    
 
\medskip\noindent\textbf{\emph{Command to run \hwcbmcv.}}
Shown below are the commands to configure \hwcbmcv in bit-level or word-level mode.
The first command using \texttt{--gen-interface} is used to generate the interface 
for the hardware modules automatically. These interface signals are required to
construct the miter as shown in column~3 of Figure~\ref{fig:hwcbmc-example}.
Note that the \texttt{<VERILOG-FILE-NAME>} can be specified as $(.v)$ or $(.sv)$
file, where $(.v)$ is an extension for Verilog files and $(.sv)$ is an extension
for SystemVerilog files. We assume that the \texttt{<MITER-FILE-NAME>} 
includes the reference model in C and implements the miter. 
Note that \hwcbmcv expects the reference model and the miter 
implementation to be C programs. The command line switch \texttt{--aig} instructs the tool 
to operate in bit-level mode. Without this option, the default operating mode
in \hwcbmcv is word-level mode. The switch \texttt{--bound} and \texttt{--unwind} 
is used to specify the unwind depth for the hardware and software transition system 
respectively. The switch \texttt{--module} specifies the name of the top level
module in the Verilog design file. \hwcbmcv also provides an option,
\texttt{--vcd} to dump counterexamples in Value Change Dump ($vcd$) format  
in case of assertion failure, which can be analyzed for debugging purpose using waveform viewer, 
such as \texttt{gtkwave}.\footnote{http://gtkwave.sourceforge.net} The switch 
\texttt{--help} shows the available command line options for using \hwcbmcv.
    
\begin{small}
\begin{verbatim}
// generate interface  
hw-cbmc <VERILOG-FILE-NAME> --module <TOP-MODULE> --gen-interface  
// bit-level mode
hw-cbmc <VERILOG-FILE-NAME> <MITER-FILE-NAME> --module <TOP-MODULE> 
--bound N --unwind M --aig --vcd <VCD-FILE-NAME> 
// word-level mode
hw-cbmc <VERILOG-FILE-NAME> <MITER-FILE-NAME> --module <TOP-MODULE> 
--bound N --unwind M --vcd <VCD-FILE-NAME> 
\end{verbatim}
\end{small}

\subsection*{Monolithic and Path-wise Approach to Equivalence Checking}

We investigated the structure of the ARM FPU and dual-path adder examples discussed the paper
to analyze the effect on runtimes 
of the monolithic and path-based equivalence checking approaches 
followed by \hwcbmcv and \verifox respectively.

We observe that the pipelined implementation of ARM FPU forces \verifox to traverse deep into a particular 
path and then backtrack to a much higher level in the symbolic tree due to 
infeasibility of the current path. This causes \verifox to throw away several 
path fragments that were earlier considered feasible while going deep in the path only to be discovered as 
infeasible much later. 
This results in the wastage of significant computation time in \verifox.
On the other hand, the dual-path adder contains a state-machine that
implements separate cases for the addition of different types of numbers. 
This allows \verifox to perform an early infeasibility check and prune most of
the irrelevant logic upfront in the symbolic execution phase using 
assumptions. On the other hand, the monolithic constraint generated by \hwcbmcv 
for the dual-path FP adder was extremely difficult to solve. In this way, 
our experiments give some insight into how the path-based symbolic execution in \verifox and the monolithic 
BMC-based approach in \hwcbmcv are sensitive to the structure of the 
original floating-point design. 

\lstdefinestyle{base}{
  language=C,
  emptylines=1,
  breaklines=true,
  basicstyle=\ttfamily\color{black},
  moredelim=**[is][\color{red}]{@}{@},
}

\begin{figure}[t]
\begin{center}
\centering \scriptsize
\begin{tabular}{|l|l|}
\hline
Miter for VERIFOX & Miter for HW-CBMC \\
\hline
\begin{lstlisting}[mathescape=true]
int miter(float f, float g) {
 roundmode          $rmode;$
 $softfloat\_uint64\_t$  $nan\_payload;$
 float          $sum\_ref, sum\_impl;$
 int ROUNDMODE;
 switch(ROUNDMODE) {
   case 0: { // ROUND TO NEAREST EVEN
     $fesetround(FE\_TONEAREST);$
     $rmode = 3;$
     $break;$
   }
   case 1: { // ROUND UP	   
     $fesetround(FE\_UPWARD);$
     $rmode = 0;$
     $break;$
   }
   case 2: { // ROUND DOWN	   
     $fesetround(FE\_DOWNWARD);$
     $rmode = 1;$
     $break;$
   }
   case 3: { // TOWARDZERO	   
     $fesetround(FE\_TOWARDZERO);$
     $rmode = 2;$
     $break;$
   }
 }
     
 // Invoke the reference model
 $sum\_ref = f + g;$
 // Invoke the ARM FPU ADD
 $nan\_payload = 0x00080000;$
 $sum\_impl = sfadd64(f, g, rmode, nan\_payload);$
 // check the output
 $assert(compareFloat(sum\_ref, sum\_impl));$
}
\end{lstlisting}
&
\begin{lstlisting}[mathescape=true]
int miter(float f, float g) {
 float $C\_result, Verilog\_result;$
 roundmode          $rmode;$
 int $ROUNDMODE;$
 // reset the design
 $add64.reset\_n = 0;$
 $set\_inputs(); next\_timeframe();$
 // pass the inputs to the RTL
 $add64.reset\_n = 1;$
 $add64.src0 = *(unsigned *)\&f; 
 add64.src1 = *(unsigned *)\&g;$
 $set\_inputs(); next\_timeframe();$
 // settings for RTL floating-point addition
 $add64.pipe\_ready = 1; add64.valid\_in = 1; add64.lane\_mask = 3;$
 switch(ROUNDMODE) {
   $case 0:$ { // ROUND TO NEAREST EVEN
     $add64.round\_mode = 0;$ 
     $rmode = 3; break;$ }
   $case 1:$ { // ROUND UP	   
     $add64.round\_mode = 1;$
     $rmode = 0; break;$ }
   $case 2:$ { // ROUND DOWN	   
     $add64.round\_mode = 2;$
     $rmode = 1; break;$ }
   $case 3:$ { // TOWARDZERO	   
     $add64.round\_mode = 3;$
     $rmode = 2; break;$ }
 }
 $set\_inputs(); next\_timeframe();$
 $next\_timeframe(); next\_timeframe();$
 $next\_timeframe();$
 // Invoke the ARM FPU ADD
 $nan\_payload = 0x00080000;$
 $C\_result$ = $sfadd64(f, g, rmode, nan\_payload);$
 // RTL result must be ready here
 $Verilog\_result=*(float *)\&add64.res;$
 // check the output
 $assert(compareFloat(C\_result, Verilog\_result));$
}
\end{lstlisting}
\\
\hline
\end{tabular}
\end{center}
\vspace{-1.7mm}
\caption{Miter for equivalence checking
         of a double precision floating-point adder from ARM}
\label{fig:harness}
\end{figure}

\subsection*{Synthesizable Constructs in \hwcbmcv}

Our Verilog front-end in \hwcbmcv support IEEE 1364.1 2005 Verilog standards.  
This includes the entire synthesizable fragment of Verilog.  The detailed list 
of synthesizable Verilog constructs supported by our Verilog front-end is
available in our website www.cprover.org/ebmc/manual/verilog\_language\_features.shtml.

\subsection*{Miter Construction for Equivalence Checking}

Figure~\ref{fig:harness} shows an example miter for checking 
equivalence of a 64-bit floating-point adder at the software 
level and RTL phase using \verifox and \hwcbmcv respectively. 

For the miter in \verifox, we provide the same floating-point 
numbers as inputs to the reference design (built inside \verifox) 
and an externally provided untimed SW implementation (in C). We 
then set the \texttt{rounding mode} of the reference model and 
the SW implementation accordingly. Subsequently, the results of 
addition from the reference model ($sum\_ref$) and the SW 
implementation ($sum\_impl$) are checked for equivalence using the function, 
\texttt{assert(compareFloat(sum\_ref, sum\_impl));}.

In a similar way, the miter in \hwcbmcv is constructed by providing the same 
floating-point numbers as input to the SW and HW RTL implementations. 
Note that the inputs are set to the HW signals in \hwcbmcv using a function 
\lstinline!set_inputs()!. Since the ARM FPU is a pipeline implementation 
with pipeline depth 4, we unwind the HW transition system up to a bound 
of 4 using the function \lstinline!next_timeframe()!. Subsequently, the 
results computed by the HW design and the C reference model are 
compared using the \lstinline!compareFloat()! function.

\subsection*{Miter for Combinational Equivalence Checking in \hwcbmcv}

\lstset{language=C,basicstyle=\ttfamily}

Figure~\ref{fig:miter} shows an example miter for checking 
combinational equivalence of a 32-bit floating-point
adder/subtractor circuit. 
\begin{figure}[h]
\begin{center}
\centering \scriptsize
\begin{tabular}{l}
\begin{lstlisting}[mathescape=true]
void miter(float f, float g) {
  // setting up the inputs to hardware FPU
  $fp\_add\_sub.f=*(unsigned *)\&f;$                        
  $fp\_add\_sub.g=*(unsigned *)\&g;$                        
  $fp\_add\_sub.isAdd=1;$                                   
  // propagates inputs of the hardware circuit    
  $set\_inputs();$                                          
  // get result from hardware circuit                         
  float $Verilog\_result=*(float *)\&fp\_add\_sub.result;$      
  // compute fp-add in Software with rounding mode RNE
  float $C\_result=add(RNE, f, g); $                            
  // compare the outputs                       
  $assert(compareFloat(C\_result, Verilog\_result));$
}                    
\end{lstlisting}
\end{tabular}
\end{center}
\vspace{-1.7mm}
\caption{Miter for combinational equivalence checking
         for a 32-bit floating-point adder/subtractor
         for the case of addition in \hwcbmcv}
\label{fig:miter}
\end{figure}
We provide the same floating-point numbers as
inputs to the reference design (in C) and the hardware implementation (in
RTL Verilog) using \lstinline!set_inputs()!.  Subsequently, we
indicate that we want to perform a floating-point addition by setting
\lstinline!isAdd=1!.  The results computed by the hardware design and the C
reference model are compared using the \lstinline!compareFloat()!  function.
Note that this is a combinational circuit, so there is no call to 
\lstinline!next_timeframe()!.

\end{document}